\documentclass[twocolumn,groupedaddress,superscriptaddress,floatfix,pra,showpacs,nofootinbib]{revtex4-2}

\usepackage{graphicx,bm,dsfont,amsmath, amsthm, amssymb,mathrsfs,setspace,hyperref,mdframed,footnote,mathtools,xcolor,natbib,soul}

\newcommand{\ket}[1]{\left| #1 \right\rangle}
\newcommand{\bra}[1]{\left\langle #1 \right|}

\begin{document}

\title{Backscatter and Spontaneous Four-Wave Mixing in Micro-Ring Resonators}

\author{Jonte R. Hance}
\email{jonte.hance@bristol.ac.uk}
\affiliation{Quantum Engineering Technology Laboratory, Department of Electrical and Electronic Engineering, University of Bristol, Woodland Road, Bristol, BS8 1UB, UK}
\author{Gary F. Sinclair}
\affiliation{Quantum Engineering Technology Laboratory, Department of Electrical and Electronic Engineering, University of Bristol, Woodland Road, Bristol, BS8 1UB, UK}
\author{John Rarity}
\affiliation{Quantum Engineering Technology Laboratory, Department of Electrical and Electronic Engineering, University of Bristol, Woodland Road, Bristol, BS8 1UB, UK}
\date{\today}

\begin{abstract}
We model backscatter for electric fields propagating through optical micro-ring resonators, as occurring both in-ring and in-coupler. These provide useful tools for modelling transmission and in-ring fields in these optical devices. We then discuss spontaneous four-wave mixing and use the models to obtain heralding efficiencies and rates. We observe a trade-off between these, which becomes more extreme as the rings become more strongly backscattered.
\end{abstract}

\maketitle
\section{Introduction}
We need sources of controlled numbers of discrete photons to create photonic circuits for quantum computing. Two established ways we can generate photons are using near-deterministic single-photon emitters (e.g. colour centres \cite{Hadden2010Strongly} and quantum dots \cite{Somaschi2016Near}) and spontaneous generation using parametric nonlinearities (e.g. four-wave mixing in optical fibre \cite{Alibart2006Photon} and silicon photonics \cite{Faruque20182Rings}).  Although parametric generation of photon pairs is probabilistic, we can mitigate this by using one half of a photon pair to herald its partner's presence \cite{Migdall2002Tailoring}.
Four-wave mixing occurs in a variety of devices, but is most conveniently produced in integrated circuits by micro-ring resonators (MRRs). These allow higher generation rates, due to resonant field enhancement \cite{Clemmen2009Continuous,Helt2010MRR,Azzini2012Ultra,Harris2014Integrated,Gentry2015RingExp,Savanier2016Photon}.

Typically, their transmission displays Lorentzian-shaped resonant peaks, reaching a minimum when the ring circumference is an integer multiple of the wavelength \cite{Bogaerts2011SiliconMRR}.
However, these devices are vulnerable to backscatter \cite{Suo2017BackscatterImpact}. 

This occurs when light couples between the forward and backward modes within the ring, either due to reflections in the coupling between bus and ring, or from the surface roughness of the waveguide. This causes a splitting of the resonance peak, reducing resonant enhancement and changing the shape of the spectral response. Furthermore, one member of a generated two-photon pair could be backscattered and lost, reducing the heralding efficiency of the photon-pair source \cite{Vernon2016NoFree}. 

However, maintaining high heralding efficiency is essential to overcoming the randomness inherent to parametric photon-pair generation.  We therefore investigate how this loss mechanism will limit performance in ring-resonator sources. 

While some, such as Li et al, have considered the effects of backscatter \cite{Li2016backscattering}, there isn't yet a full analytic model for its effects on field propagation through a ring. Here we construct this unified analytic model for backscatter in both the ring and the coupler, and we apply it to spontaneous four-wave mixing in an MRR. This allows us to analyse the trade-offs between heralding rate and heralding efficiency.

\begin{figure}
    \centering 
    \includegraphics[width=\linewidth]{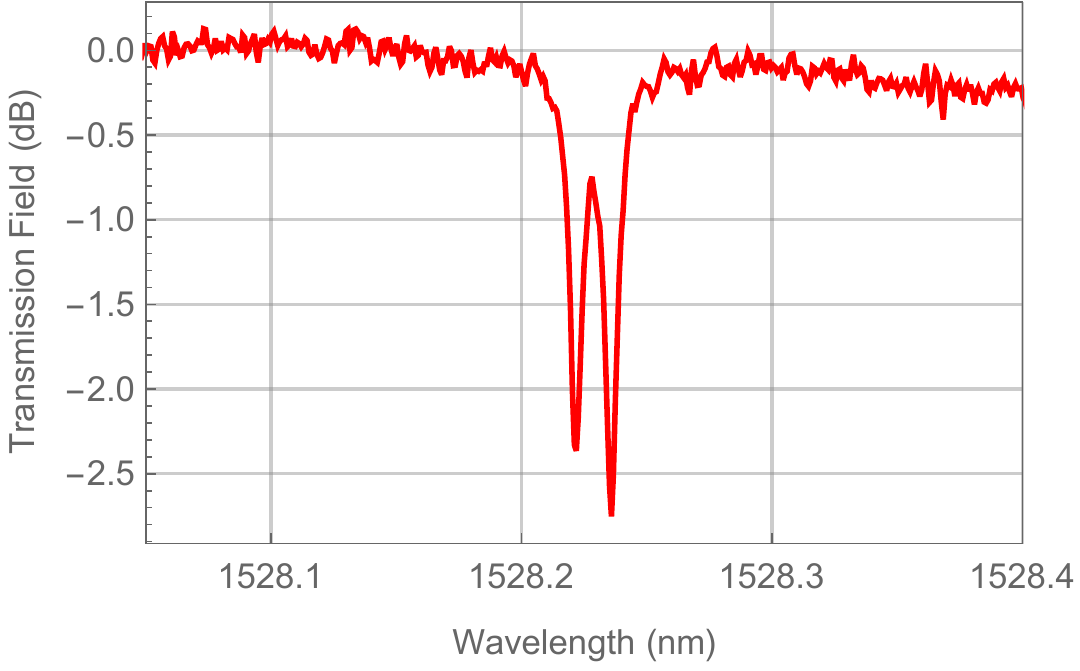}
    \caption{Transmission power obtained experimentally from an Micro-Ring Resonator, showing normalised output power ($y$-axis) plotted against pump wavelength ($x$-axis, in $nm$), for a $5\mu m$ ring. It shows the a fundamental resonance dip split by  backscatter effects.}
    \label{fig:expgraph}
\end{figure}

While previous studies have looked at how the heralding efficiency is limited by design parameters \cite{Vernon2016NoFree} and material properties such as cross two-photon absorption in silicon \cite{Husko2013Multi, Sinclair2019Temperature}, this is the first study that investigate the role of backscatter on this, which can result from fabrication imperfections or non-optimal design. Therefore, this research will be useful to anyone designing MRRs for generation of photon pairs. Fig. \ref{fig:expgraph} presents a typical experimental transmission spectrum  from a Micro-Ring Resonator showing the characteristic asymmetric split peaks, which any analytic model of backscatter must be able to explain in order to fully model the effect.

\section{Modelling Backscatter}
\subsection{Matrix Formalism}

\begin{figure}
    \centering 
    \includegraphics[width=\linewidth]{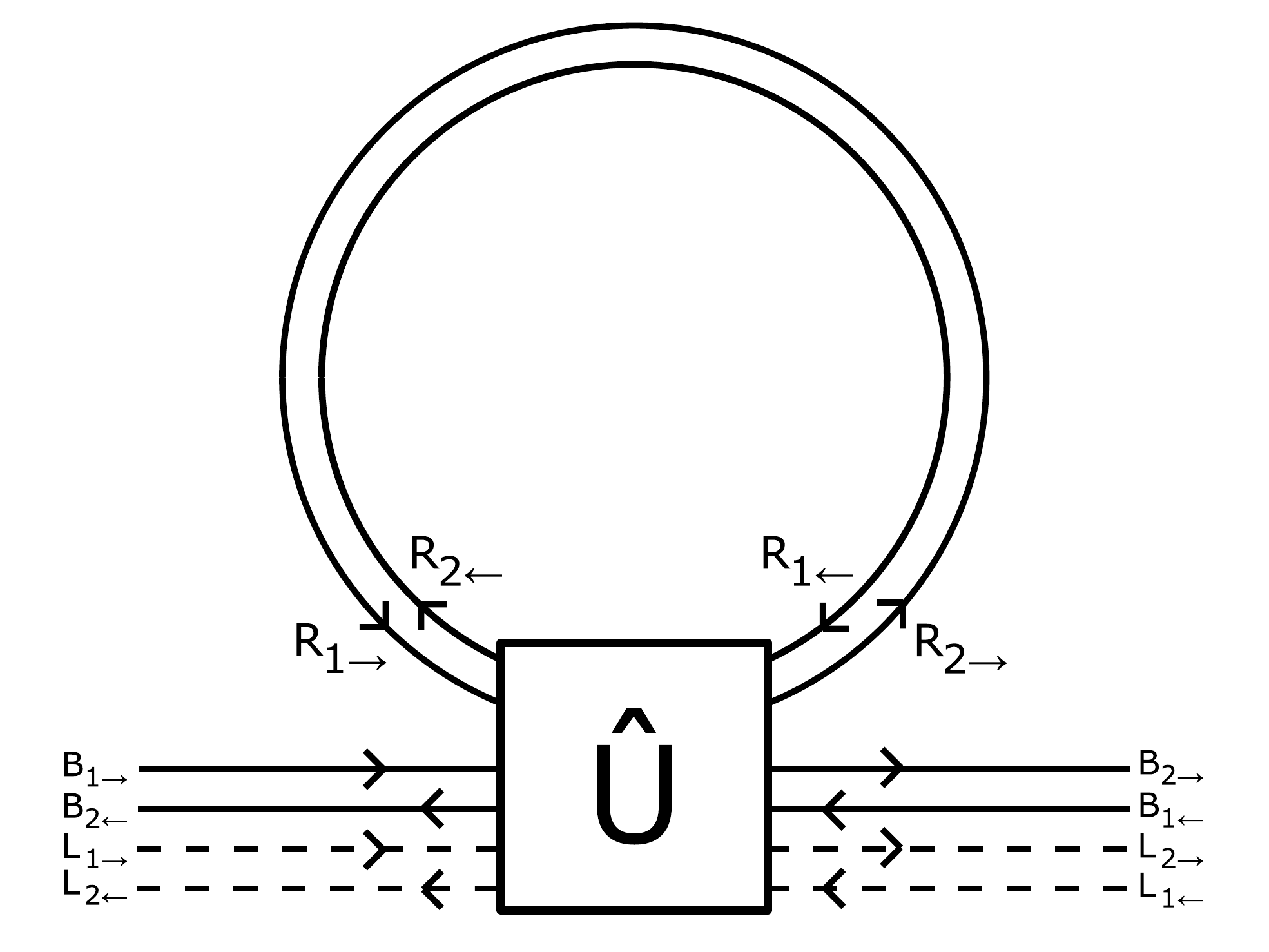}
    \caption{Our matrix model of an MRR, with three two-mode vectors ($\textbf{R}$, $\textbf{B}$ and $\textbf{L}$), each with a forward and a backward component ($\rightarrow$ and $\leftarrow$), all linked by the $6\times6$ unitary matrix $\textbf{U}$. As all interaction occurs in this matrix, $\textbf{R$_1$}=\textbf{R$_2$}$.}
    \label{fig:matring}
\end{figure}

We model the system as a scattering matrix over six modes, representing the forward and backward fields in the ring, bus and loss channel, as per Fig. \ref{fig:matring}.

\begin{equation}
\begin{bmatrix}
R_{2\rightarrow}\\
R_{2\leftarrow}\\
B_{2\rightarrow}\\
B_{2\leftarrow}\\
L_{2\rightarrow}\\
L_{2\leftarrow}\\
\end{bmatrix}
=\textbf{U}
\begin{bmatrix}
R_{1\rightarrow}\\
R_{1\leftarrow}\\
B_{1\rightarrow}\\
B_{1\leftarrow}\\
L_{1\rightarrow}\\
L_{1\leftarrow}\\
\end{bmatrix}
\end{equation}
where R, B and L correspond to ring, bus and loss modes, $\rightarrow$ and $\leftarrow$ to forward- and backward-travelling fields, and 1 and 2 to entering and leaving the scattering matrix. By modelling loss via coupling to a fictional mode, we conserve unitarity, and so the commutation relations, making the model suitable for later adaption for quantum analysis.

Note we model $L_{1\rightarrow}=L_{1\leftarrow}=0$ (as noise input from the vacuum will be far lower than the powers we are considering); and, for consistency (as we model all interaction as taking place within $\textbf{U}$) $R_{1\rightarrow}=R_{2\rightarrow}, \; R_{1\leftarrow}=R_{2\leftarrow}$.

Our interaction can be thought of as consisting of a number of compiled smaller interactions between two modes. Therefore, the $6\times6$ matrices representing each process $\textbf{Cpl}_{BR}$, $\textbf{Loss$_{RL}$}$, $\textbf{Back$_R$}$ and $\textbf{Back$_C$}$ (bus-ring coupling, ring-loss coupling, and backscatter, modelled in-ring and in-coupler respectively) are
\begin{align}
        \textbf{Cpl}_{BR}=& 
        \textbf{T}_{B_{\rightarrow}R_{\rightarrow}}+\textbf{T}_{B_{\leftarrow}R_{\leftarrow}}+
        \mathds{1}_L\\
        \textbf{Loss}_{RL}=& 
        \textbf{A}_{R_{\rightarrow}L_{\rightarrow}}+\textbf{A}_{R_{\leftarrow}L_{\leftarrow}}+
        \mathds{1}_B\\
        \textbf{Back}_{R}=& 
        \textbf{C}^R_{R_{\rightarrow}R_{\leftarrow}}+
        \mathds{1}_{L,B}\\
        \textbf{Back}_{C}=& 
        \textbf{C}^C_{R_{\rightarrow}R_{\leftarrow}}+\textbf{C}^C_{B_{\rightarrow}B_{\leftarrow}}+\textbf{C}^C_{L_{\rightarrow}L_{\leftarrow}}
\end{align}
where $\mathds{1}$ is the identity matrix on a given mode (meaning the process matrix doesn't alter that mode), and sub-matrices (on two modes, here $a$ and $b$), are
\begin{align}
    \textbf{A}_{ab}=
    \begin{bmatrix}
    e^{-i\theta(\lambda)}\alpha & e^{i\theta(\lambda)}\sqrt{1-\alpha^2}\\
    -e^{-i\theta(\lambda)}\sqrt{1-\alpha^2}& e^{i\theta(\lambda)}\alpha
    \end{bmatrix}_{ab}\\
    \textbf{C}^R_{ab}=
    \begin{bmatrix}
    |\xi| & e^{i\zeta}\sqrt{1-|\xi|^2}\\
    -e^{-i\zeta}\sqrt{1-|\xi|^2}& |\xi|
    \end{bmatrix}_{ab}\\
    \textbf{C}^C_{ab}=
    \begin{bmatrix}
    e^{-i\zeta/2}|\xi|  &  e^{i\zeta/2}\sqrt{1-|\xi|^2}\\
    -e^{-i\zeta/2}\sqrt{1-|\xi|^2}  & e^{i\zeta/2}|\xi| 
    \end{bmatrix}_{ab}\\
    \textbf{T}_{ab}=
    \begin{bmatrix}
    e^{-2i\phi}|t|  & \sqrt{1-|t|^2}\\
    -e^{-2i\phi}\sqrt{1-|t|^2}  & |t|
    \end{bmatrix}_{ab}
\end{align}
where $\alpha$ is the loss coefficient (one with no loss in the ring, nil with complete loss), $\xi$ the backscatter coefficient (one when no backscatter, zero when entirely backscattered), and $t$ the coupling coefficient (one with no ring-bus coupling, zero with total coupling), and $\zeta$ and $\phi$ the respective phases on backscatter and ring-bus coupling. 

Note backscatter modelled in-ring and in-coupler (in $\textbf{C}^R_{ab}$ and $\textbf{C}^C_{ab}$) differ by backscatter in-ring having all phase occurring on the backward component, and backscatter in-coupler having opposite phase for the forward and backward components. Both have the same phase difference, $\zeta$, between forward and backward components.

$\theta$ is the phase the field accrues over one trip around the ring, 
\begin{equation}
    \theta(\lambda) = \frac{4 \pi^2 n_e r}{\lambda}+\tau
\end{equation}
where $n_e$ is the effective refractive index, $r$ is the ring radius, $\tau$ is any phase offset caused by the loss $\alpha$, and $\lambda$ is the field wavelength.

For instance, $\textbf{Back}_R$ is
\begin{equation}
    \begin{bmatrix}
    |\xi|& e^{i\zeta}\sqrt{1-|\xi|^2} & 0 & 0 & 0 & 0\\
    -e^{-i\zeta}\sqrt{1-|\xi|^2}& |\xi| & 0 & 0 & 0 & 0\\
    0 & 0 & 1 & 0 & 0 & 0\\
    0 & 0 & 0 & 1 & 0 & 0\\
    0 & 0 & 0 & 0 & 1 & 0\\
    0 & 0 & 0 & 0 & 0 & 1
    \end{bmatrix}
\end{equation}

From this, the two unitary matrices, representing the entire interaction (for in-ring and in-coupler backscatter respectively), are
\begin{align}
        \textbf{U}_R=\textbf{Cpl}_{BR}\cdot\textbf{Back}_{R}\cdot\textbf{Loss}_{RL}\\
         \textbf{U}_C=\textbf{Cpl}_{BR}\cdot\textbf{Back}_{C}\cdot\textbf{Loss}_{RL}
\end{align}

Note, the above ordering doesn't matter, as these process matrices commute (up to arbitrary powers of $-1$).

\subsection{Transmission}

Now we can obtain the transmission, to compare with observations (e.g. Fig. \ref{fig:expgraph}) by, (where row-column subscript notation picks out $2\times2$ sub-matrices),
\begin{equation}
\begin{split}
    \begin{bmatrix}
    B_{2\rightarrow}\\
    B_{2\leftarrow}
    \end{bmatrix}
    =\textbf{U}_{BR} \begin{bmatrix}
    R_{1\rightarrow}\\
    R_{1\leftarrow}
    \end{bmatrix}
    + \textbf{U}_{BB} \begin{bmatrix}
    B_{1\rightarrow}\\
    B_{1\leftarrow}
    \end{bmatrix}&\\
    =
    \Bigg(
    \frac{\textbf{U}_{BR}\textbf{U}_{RB}}{\mathds{1}_2-
    \textbf{U}_{RR}} + 
    \textbf{U}_{BB}
    \Bigg)
    \begin{bmatrix}
    B_{1\rightarrow}\\
    B_{1\leftarrow}
    \end{bmatrix}&
\end{split}
\end{equation}

This gives the transmission, $B_{2\rightarrow}$, for in-ring backscatter and in-coupler backscatter,
\begin{align}
    B^R_{2\rightarrow}=\frac{1}{t}-\frac{(t^{-1}+\alpha|\xi| e^{-i\theta})(|t|^2-1)}{1+2t\alpha e^{-i\theta}|\xi|+(t\alpha e^{-i\theta})^2}\\
    B^C_{2\rightarrow}=\frac{\xi}{t}-\frac{(\xi-t\alpha e^{-i\theta})(|t|-|t|^{-1})}{(t\alpha e^{-i\theta})^2-t\alpha e^{-i\theta}(\xi+\xi^*)+1}
\end{align}
when $B_{1\rightarrow}$ is normalised to 1, and $B_{1\leftarrow}$ is zero.

\begin{figure*}
 \centering 
    \includegraphics[width=\linewidth]{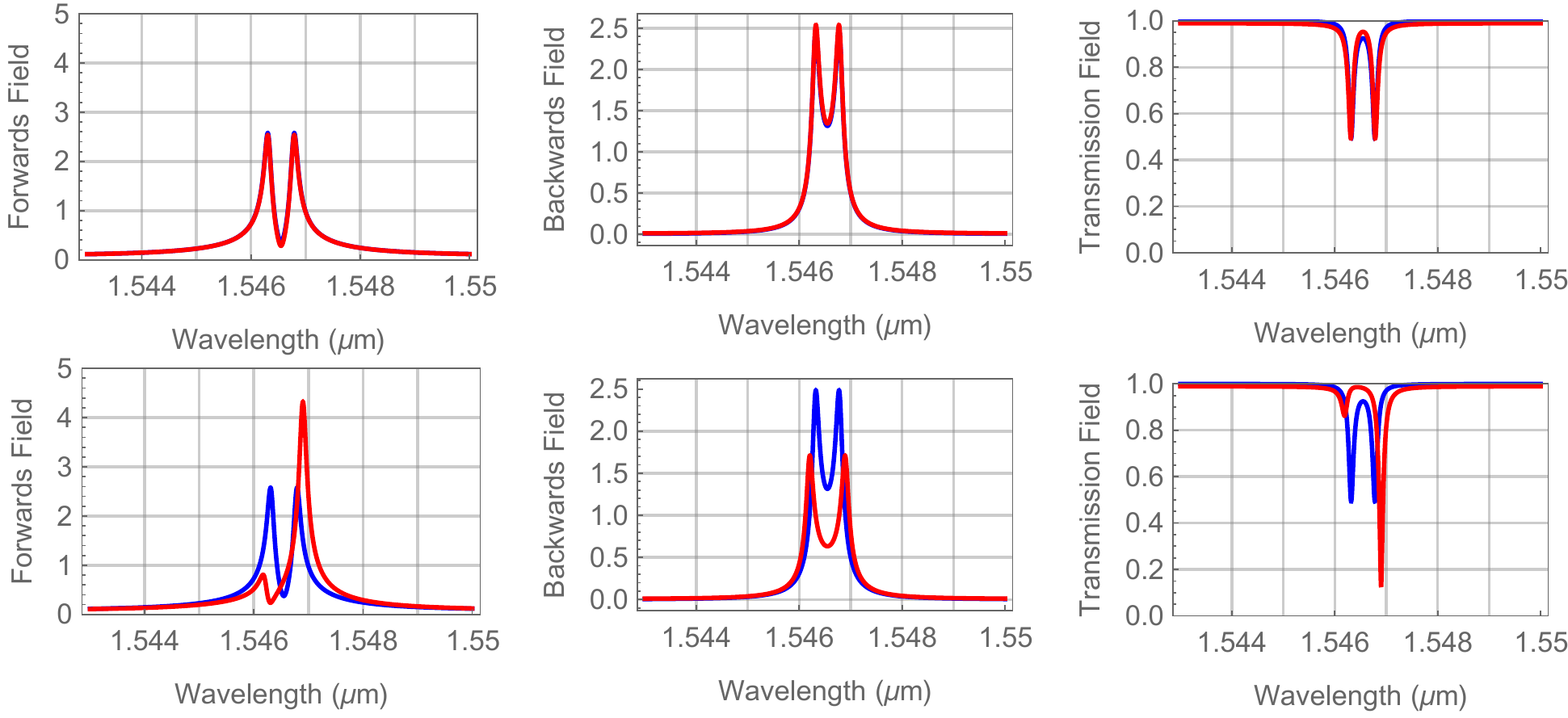}
    \caption{Absolute values of both forward (left column) and backward (middle) ring fields, and transmission (right), for both in-ring (blue) and in-coupler (red) models. The $x$-axis is wavelength in $\mu m$, and the $y$-axis is field magnitude normalised against input field, with ring-bus coupling coefficient $t$ and loss coefficient $\alpha$ of magnitude 0.98 and ring-bus coupling phase $\phi$ of 0, backscatter coefficient $\xi$ of magnitude 0.99 and phase $\zeta$ of 0 (top row) and $\pi$/10 (bottom), for ring radius $r$ of $15\mu m$.}
    \label{fig:mfieldgraphphase}
\end{figure*}

(Note, Biasi et al also give an analytical model for backscattering (albeit in microdisk resonators) \cite{Biasi2018ModeCoupling} - however, their model is based on the time-differential of the field, whereas our model is time-independent, based on a scattering matrix. As future work, we would be interested in seeing their model applied to the four-wave mixing analysis presented in Section \ref{section:Heralding}.)

\subsection{Field In-Ring}
Due to how the scattering matrices were composed, the ring field values they give, $\textbf{R}$, are those after a full ring round-trip. However, we need the average power in the ring to work out the photon pairs generated by spontaneous four-wave mixing - and so the field halfway through the ring. Therefore, we need to reorder the full interaction matrices, giving
\begin{align}
        \textbf{U}_R=&\textbf{Loss}_{RL}^\frac{1}{2}\cdot\textbf{Back}_{R}^\frac{1}{2}\cdot\textbf{Cpl}_{BR}\cdot\textbf{Back}_{R}^\frac{1}{2}\cdot\textbf{Loss}_{RL}^\frac{1}{2}\\
         \textbf{U}_C=&\textbf{Loss}_{RL}^\frac{1}{2}\cdot\textbf{Cpl}_{BR}\cdot\textbf{Back}_{C}\cdot\textbf{Loss}_{RL}^\frac{1}{2}
         \label{eq:inringmatr}
\end{align}

These give the same transmission as before, but now also the average in-ring fields. Focusing on the sub-matrices, considering the previous method, the relations required are
\begin{equation}
\begin{bmatrix}
    R_{\rightarrow}\\
    R_{\leftarrow}
    \end{bmatrix}
    =\frac{\textbf{U}_{RB}}{\mathds{1}_2-
    \textbf{U}_{RR}}
    \begin{bmatrix}
    B_{1\rightarrow}\\
    B_{1\leftarrow}
    \end{bmatrix}
\end{equation}

For both in-ring and in-coupler backscatter, this gives equations for the forward and backward fields. Unfortunately, these expressions are far more complicated than that for transmission - their derivation, and graphs of produced responses, are far more informative than their actual formulae.

From this, we get forward and backward pump power in-ring as
\begin{equation}
\begin{bmatrix}
    P_{p\rightarrow}\\
    P_{p\leftarrow}
    \end{bmatrix}
    =
    \Bigg(\frac{\textbf{U}_{RB}}{\mathds{1}_2-
    \textbf{U}_{RR}}
     \begin{bmatrix}
    B_{1\rightarrow}\\
    B_{1\leftarrow}
    \end{bmatrix}
    \Bigg)^2
\end{equation}

\subsection{Comparison of In-Ring and In-Coupler Scattering}

Fig. \ref{fig:mfieldgraphphase} shows the above models are the same when the backscatter phase is nil. However, when a phase is applied, the in-coupler model peaks show asymmetry, while the in-ring model shows none.

The in-ring model's lack of asymmetric split peaks is un-physical, given we see asymmetric peak in transmission spectra of MRRs (e.g. \cite{Li2016backscattering}, and Fig. \ref{fig:expgraph}). This symmetry could be as we don't allow the backscatter coefficient, $\xi$, to have a phase for the in-ring model, given, unlike for in-coupler, it makes no sense for backscatter to apply a phase to the forward-coupled component when scattering occurs in-ring.

To replicate experimental data, it makes sense to combine the models, to have backscatter within both ring and coupler. This makes sense, as backscatter has been associated with different things in each case: for in-ring, waveguide-roughness; and for in-coupler, mode-mismatch between the straight and curved coupling regions, and increased roughness-induced backscatter due to higher field intensity in the coupling region.

However, both models give the same amount of splitting for a given backscatter coefficient, and are equivalent when phase is nil. This means it makes sense to just use the backscatter in-coupler model, and proportionally reduce the phase on the backscatter coefficient. This removes unnecessary degrees of freedom, and allows us to more easily equate coupling constants with observables.

\section{Issues with the Backscatter Models}

\subsection{Issues with Point-Coupling Model}
An issue with the models is that they all treat coupling as occurring at a fixed point, rather than continuously. This was been raised as an issue in Li et al \cite{Li2016backscattering}. However, this poses less of an issue than initially expected.

This is as, when backscatter is modelled in-ring, coupling at one point is equivalent to applying the $N^{th}$ root of the coupling $N$ times distributed around the ring. We see this when, for determining the in-ring field in the in-ring matrix model, the square root of the backscatter matrix is applied twice - once before and once after the field value is taken. At its limit, this single point-backscatter model is equivalent to applying the differentiated backscatter matrix at every point along the circumference - all this requires is the backscatter being phase-coherent. This seems rational, and necessary for any adequate modelling of the system, so we accept it, allowing us to treat in-ring point coupling as equivalent to uniform coupling at all points.

This can also be applied to the ring-bus coupling, across the coupling length (weighted by ring-bus distance). Again, this shows single-point coupling is a simpler, but just as apt, model as continuous coupling. However, it relies on the assumption that group velocity is the same in both ring and bus, which isn't necessarily so.

\subsection{Assumption of Equal Group Velocity}
Both models are in the frequency domain, without time-dependence, so the wave velocity doesn't have to be factored in. However, to calculate spontaneous four-wave mixing, the field derivative with respect to time has to be taken, which means moving from the frequency to time domain \cite{Sipe2015LossyRing,Alsing2017Loss}.
In that, the photon would be treated as a fixed-width Gaussian wave-packet travelling through the waveguide.

While this works when bus and ring waveguides have equal group velocities, it doesn't when they differ. Then, the wave-packet travels at different rates on either side of the coupler, meaning the non-dispersive assumption, that the rate of change of the wave-packet width remains constant, can't be upheld. This is compounded by the fact that the group index, and so the speed of light in the medium, itself varies with the field intensity, making the situation even harder to model. 

However, based on our situation, this can be neglected, as the wavelengths typical for integrated photonics ($\sim 1.55\mu m$) are at the zero-dispersion point for our typical waveguide structure (silicon-on-silica) \cite{Dekker2007Ultrafast}. Therefore, as dispersion will be nil regardless of the group velocity here, the effect can be neglected, and the model applied to this situation.

\section{Heralding 
%Purity and
Efficiency}\label{section:Heralding}

\begin{figure*}
 \centering 
    \includegraphics[width=\linewidth]{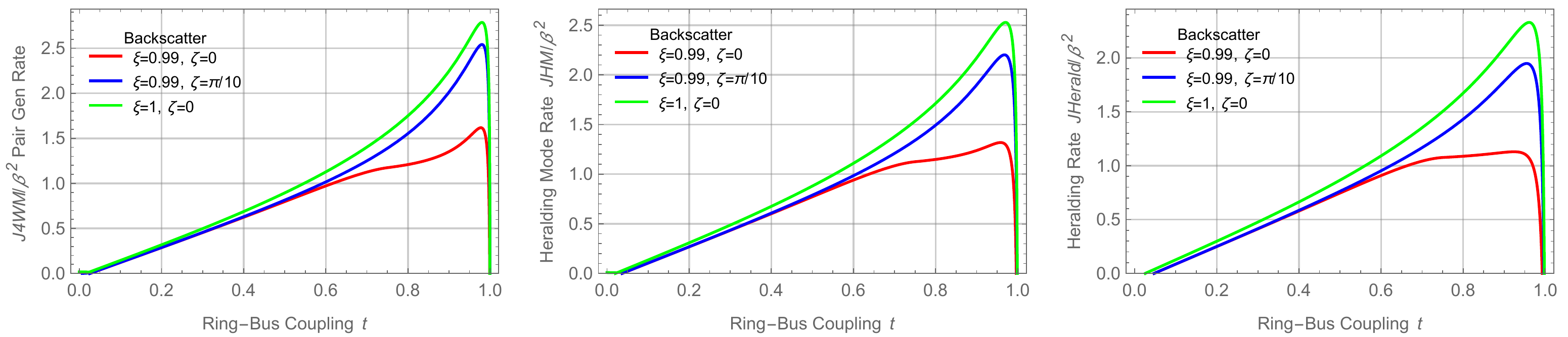}
    \caption{The maximal reduced pair generation rate $J_{4WM}/\beta^2$, heralding mode rate $J_{HM}/\beta^2$, and heralding rate $J_{Herald}/\beta^2$. The backscatter coefficient $\xi$ has magnitude 0.99 and phase $\zeta$ of 0 (red), magnitude 0.99 and phase $\pi/20$ (blue), and magnitude 1 (no backscatter, green). The $x$-axis is the ring-bus coupling coefficient $t$, and the $y$-axis is the logarithm in base-10 of the rate, with loss coefficient $\alpha$ of 0.98, for normalised input field $B_{1\rightarrow}$.}
    \label{fig:maxrates}
\end{figure*}

\subsection{Paired Photon Generation Rate}

We now want to take this model for backscatter in ring resonators, and obtain the heralding rate and efficiency. To do this, we first need the photon-pair generation rate.

Wang et al's work on spontaneous four-wave mixing \cite{Wang2001S4WMFibre}, adapted to an MRR (setting the length $L$ to one round-trip around the ring, $2\pi r$), gives the average photon-pair generation rate as
\begin{equation}
\begin{split}
    J_{4WM}=&\bra{vac}\hat{n}_s(K,2\pi nr/c)\ket{vac}=\Big|\beta 
    \begin{bmatrix}
    P_{p\rightarrow}\\
    P_{p\leftarrow}
    \end{bmatrix}\Big|^2\\
    =&\Big|\frac{3\pi^2\epsilon_0c\chi^{(3)}r}{2n_p^2\lambda A_{eff}}
    \Bigg(\frac{\textbf{U}_{RB}}{\mathds{1}_2-
    \textbf{U}_{RR}}
     \begin{bmatrix}
    B_{1\rightarrow}\\
    B_{1\leftarrow}
    \end{bmatrix}\Bigg)^2\Big|^2
\end{split}
\end{equation}
where
\begin{equation}
    \beta=\frac{3\pi^2\epsilon_0c\chi^{(3)}r}{2n_p^2\lambda A_{eff}}
\end{equation}

$n_p$ is the effective index for the pump wavelength, $A_{eff}$ is the cross-sectional area of the waveguide, and $\chi^{(3)}$ is the third-order nonlinear response of silicon. In the backscatter-free limit, this is identical to the result Vernon et al obtain \cite{Vernon2016NoFree}. As shown by Wang et al \cite{Wang2001S4WMFibre}, these photons are always generated in pairs, so we don't need to consider its effects on efficiency.

\subsection{Ring Effects on Generated Photons}

We now want the proportion of photons emitted from the device, out of those created into a given mode. As photon number proportions behave similarly to light field intensity, we can, similarly to when deriving the field strength in the ring, set
\begin{equation}
\begin{bmatrix}
\sqrt{Pr_{R\rightarrow}}\\
\sqrt{Pr_{R\leftarrow}}\\
\sqrt{Pr_{B\rightarrow}}\\
\sqrt{Pr_{B\leftarrow}}\\
\sqrt{Pr_{L\rightarrow}}\\
\sqrt{Pr_{L\leftarrow}}\\
\end{bmatrix}
=\textbf{U}
\begin{bmatrix}
\sqrt{Pr_{R\rightarrow}+1}\\
\sqrt{Pr_{R\leftarrow}+q^2}\\
0\\
0\\
0\\
0\\
\end{bmatrix}
\end{equation}

Here, R, B and L correspond to ring, bus and loss modes, $\rightarrow$ and $\leftarrow$ to forward- and backward-travelling fields, and the unitary $\textbf{U}$ is determined for the in-ring and in-coupler models by Eq.\ref{eq:inringmatr}.
Factor $q$ is the ratio of the power backwards (from the backward-travelling pump field) to that forwards,
\begin{equation}
  q=\frac{|P_{p\leftarrow}|}{|P_{p\rightarrow}|}
\end{equation}

For this model, we assume the various coupling and scattering parameters remain the same for all modes - that they don't vary with frequency. This is valid for our narrow-band assumption (that signal and idler are both within a $\mathcal{O}(1\%)$ band around the pump  frequency), and means the parameters in $\textbf{U}$ are identical to those earlier, with the exception of $ \textbf{C}^C_{ab}$, which changes to
\begin{equation}
    \textbf{C}^C_{ab}=
    \begin{bmatrix}
    e^{-i\zeta/2}|\xi|  &  0\\
    -e^{-i\zeta/2}\sqrt{1-|\xi|^2}  &  1
    \end{bmatrix}
    \begin{bmatrix}
    a\\
    b
    \end{bmatrix}
\end{equation}
as photons coming from the backward into the forward mode won't be part of a coherent pair.

By rearranging, we get
\begin{equation}
\begin{split}
    \begin{bmatrix}
    Pr_{B\rightarrow}\\
    Pr_{B\leftarrow}
    \end{bmatrix}
    =&\Bigg(\textbf{U}_{BR}\begin{bmatrix}
    \sqrt{Pr_{R\rightarrow}+1}\\
    \sqrt{Pr_{R\leftarrow}+q^2}
    \end{bmatrix}\Bigg)^2\\
    =&\Bigg(
   \textbf{U}_{BR}\sqrt{\frac{\textbf{U}_{RR}^2}{\mathds{1}_2-
    \textbf{U}_{RR}^2}+\mathds{1}_2}\Bigg)^2
    \begin{bmatrix}
    1\\
    q^2
    \end{bmatrix}
    \\
    =&
   \frac{\textbf{U}_{BR}^2}{\mathds{1}_2-
    \textbf{U}_{RR}^2}
    \begin{bmatrix}
    1\\
    q^2
    \end{bmatrix}
    \end{split}
\end{equation}

This gives the proportion of photons emitted, in a given mode, to those created. This is maximised when on resonance, which occurs when $\theta$ is an integer multiple of $\pi$, minus the phases on any elements (e.g. minus $\zeta/2$ if asymmetrically split). Assuming the wavelengths for signal and idler obey both this resonance matching, and the four-wave mixing conditions from pump frequency, we can assume that this maximum is constant (as Vernon et al do), if we neglect effects of spectral correlation \cite{Vernon2016NoFree}.

\subsection{Heralding Rate}

\begin{figure}
 \centering 
    \includegraphics[width=\linewidth]{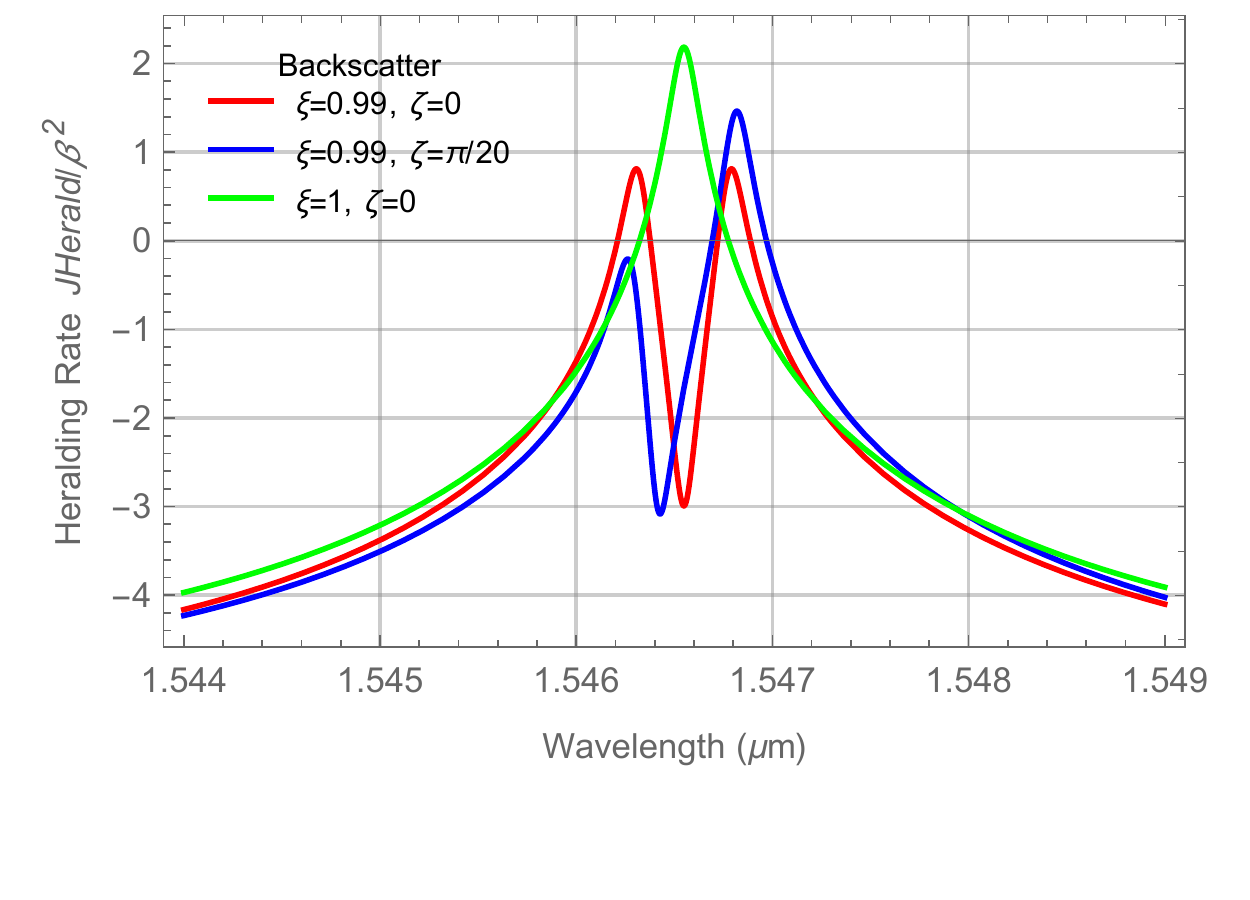}
    \caption{The heralding rate $J_{Herald}/\beta^2$, given for backscatter coefficient $\xi$ of magnitude 0.99 and phase $\zeta$ of 0 (red), magnitude 0.99 and phase $\pi/20$ (blue), and magnitude 1 (no backscatter, green). The $x$-axis is the pump wavelength in $\mu m$, and the $y$-axis is the logarithm (base-10) of the rate, with ring-bus coupling coefficient $t$ and loss coefficient $\alpha$ of magnitude 0.98 and ring-bus coupling phase $\phi$ of 0, for for normalised input field $B_{1\rightarrow}$.}
    \label{fig:heraldrates}
\end{figure}

By taking the maximal output proportion, $Pr_{B\rightarrow}$, and squaring it, we get the proportion of signal-idler pairs where both photons are emitted.
Multiplying this by the average photon-pair number generated, $J_{4WM}$, gives the average photon-pair rate - the heralding rate, $J_{Herald}$
\begin{equation}
    J_{Herald}=J_{4WM}Pr_{B\rightarrow}^2
\end{equation}
as shown in Fig.s \ref{fig:maxrates} and \ref{fig:heraldrates}. This shows, even at its peak, a relatively minor amount of backscatter reduces the heralding rate to nearly a tenth of its backscatter-free value. Alongside this, we define the rate of photons being in the Heralding Mode, $J_{HM}$, by
\begin{equation}
    J_{HM}=J_{4WM}Pr_{B\rightarrow}
\end{equation}
for which the maximal rate is again shown in Fig. \ref{fig:maxrates}, again showing a large drop (here a reduction to one-third of the original value) just to backscatter.

\subsection{Heralding Efficiency}

\begin{figure}
 \centering 
    \includegraphics[width=\linewidth]{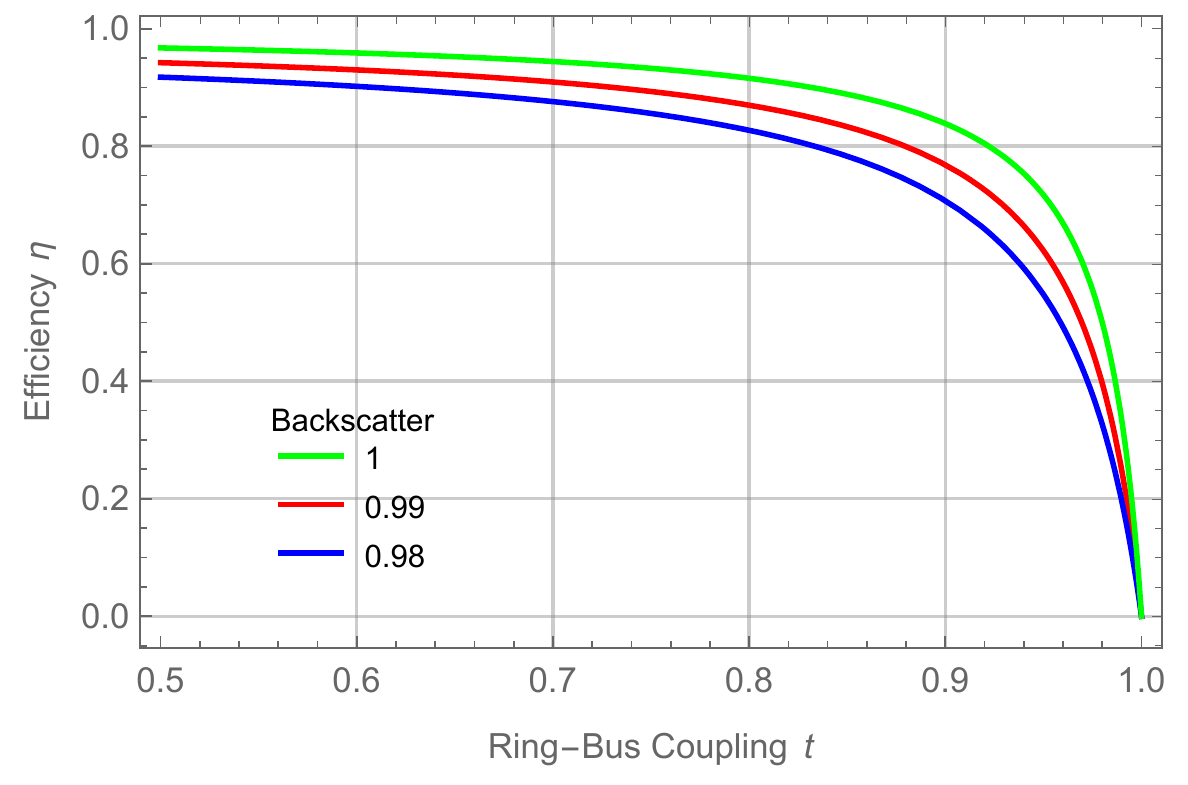}
    \caption{The heralding efficiency $\eta$, given for backscatter coefficient $\xi$ of magnitude 0.98 (blue), 0.99 (red), and 1 (no backscatter, green). The $x$-axis is the ring-bus coupling coefficient $t$, and the $y$-axis is the normalised proportion, with loss coefficient $\alpha$ of 0.98. At critical coupling ($t=\alpha=0.98$), efficiency is below 0.6 without backscatter, and below 0.3 with backscatter of coefficient 0.98, showing the effect of backscatter on heralding efficiency.}
    \label{fig:efficiency}
\end{figure}

From Vernon et al, we define the heralding efficiency, $\eta$, as the ratio between there being a heralded output photon, and a heralding photon being emitted:
\begin{equation}
    \begin{split}
        \eta = \frac{J_{Herald}}{J_{HM}}=Pr_{B\rightarrow}
    \end{split}
\end{equation}

The maximal output proportion and efficiency are the same.
This efficiency, and so the maximal output proportion, is shown for different ring-bus coupling rates in Fig. \ref{fig:efficiency}. This shows that this efficiency drops heavily as the coupling between ring and bus is reduced to nil (as $t$ goes to unity), with it going below 0.4 as we reach critical coupling (when in-ring loss and bus-ring coupling are equal, and so typical resonant peaks for bus transmission are deepest). This shows, despite critical coupling being when the most power goes into the ring when on-resonance, it most probably isn't the optimal coupling for paired photon generation. To investigate this further, we need to obtain a relationship directly between heralding rate and efficiency.

\subsection{Relationship between Rate and Efficiency}

\begin{figure}
 \centering 
    \includegraphics[width=\linewidth]{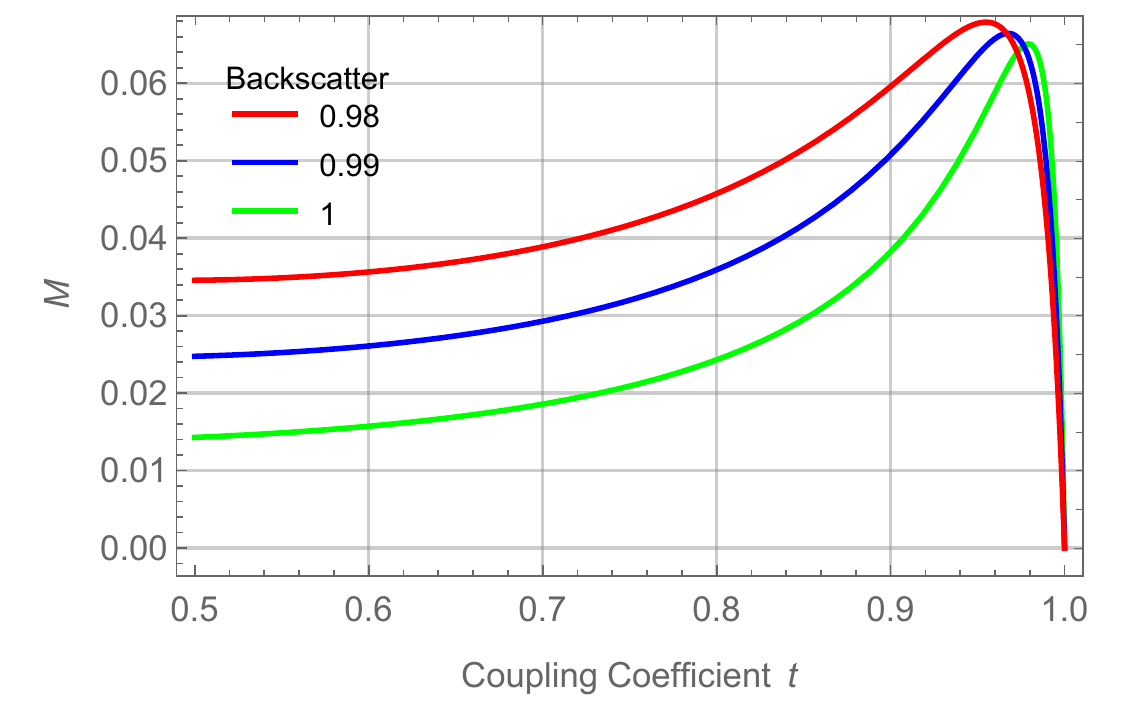}
    \caption{The value of Vernon et al's parameter $M$, as taken from Eq.\ref{VernonMEqn}, given for backscatter coefficient $\xi$ of magnitude 0.98 (blue), 0.99 (red), and 1 (no backscatter, green). The $x$-axis is the ring-bus coupling coefficient $t$, and the $y$-axis is the normalised proportion, with loss coefficient $\alpha$ of 0.98, for ring radius $r$ of $15\mu m$. Note, unlike in Vernon et al's paper, here, this is not a constant with respect to $t$.}
    \label{fig:VernonMovert}
\end{figure}

We want to define the relationship between heralding rate, $J_{Herald}$, and heralding efficiency, $\eta$. From the relationship between maximal output proportion and efficiency above,
\begin{equation}
    \begin{split}
    J_{Herald} &= J_{4WM\rightarrow}\cdot\eta^2\\
        &=\Big(\frac{3\pi^2\epsilon_0c\chi^{(3)}r}{2n_p^2\lambda A_{eff}}\Big)^2|Pr_{B\rightarrow}P_{p\rightarrow}|^2 
    \end{split}
\end{equation}

As the sub-matrices commute with one another,
\begin{equation}
\begin{split}
  \begin{bmatrix}
    Pr_{B\rightarrow}\\
    Pr_{B\leftarrow}
    \end{bmatrix}
    \begin{bmatrix}
    P_{p\rightarrow}\\
    P_{p\leftarrow}
    \end{bmatrix}=&\\
    \Bigg(\Big(  \frac{\textbf{U}_{BR}^2}{\mathds{1}_2-
    \textbf{U}_{RR}^2}\Big)&
    \Big(\frac{\textbf{U}_{RB}}{\mathds{1}_2-
    \textbf{U}_{RR}}\Big)
     \begin{bmatrix}
    B_{1\rightarrow}\\
    B_{1\leftarrow}
    \end{bmatrix}\Bigg)^2\\
    =&\frac{\eta^2(1-\eta)}{M}
     \begin{bmatrix}
    B_{1\rightarrow}^2\\
    B_{1\leftarrow}^2
    \end{bmatrix}
    \end{split}
\end{equation}
where
\begin{equation}
M=\frac{\textbf{U}_{BR}^2\Big(\textbf{1}_{2}-\textbf{U}_{RR}^2-\textbf{U}_{BR}^2\Big)}{\Big(\textbf{1}_{2}-\textbf{U}_{RR}^2\Big)^2\Big(\textbf{1}_{2}+\textbf{U}_{RR}\Big)^2}
\label{VernonMEqn}
\end{equation}

Therefore, we can still write the heralding rate as
\begin{equation}
    J_{Herald}
        =\beta^2\frac{\eta^4(1-\eta)^2}{M^2}B_{1\rightarrow}^4
\end{equation}

This is the relationship shown by Vernon et al \cite{Vernon2016NoFree}. However, as opposed to their conclusion, Fig. \ref{fig:VernonMovert} shows that this $M$ does not remain constant across all all possible ring-bus coupling strengths.

This leads us to ask if we could write the heralding rate $J_{Herald}$ as some function of the heralding efficiency $\eta$, multiplied by some constant of the ring-bus coupling $t$.

\begin{figure}
 \centering 
    \includegraphics[width=\linewidth]{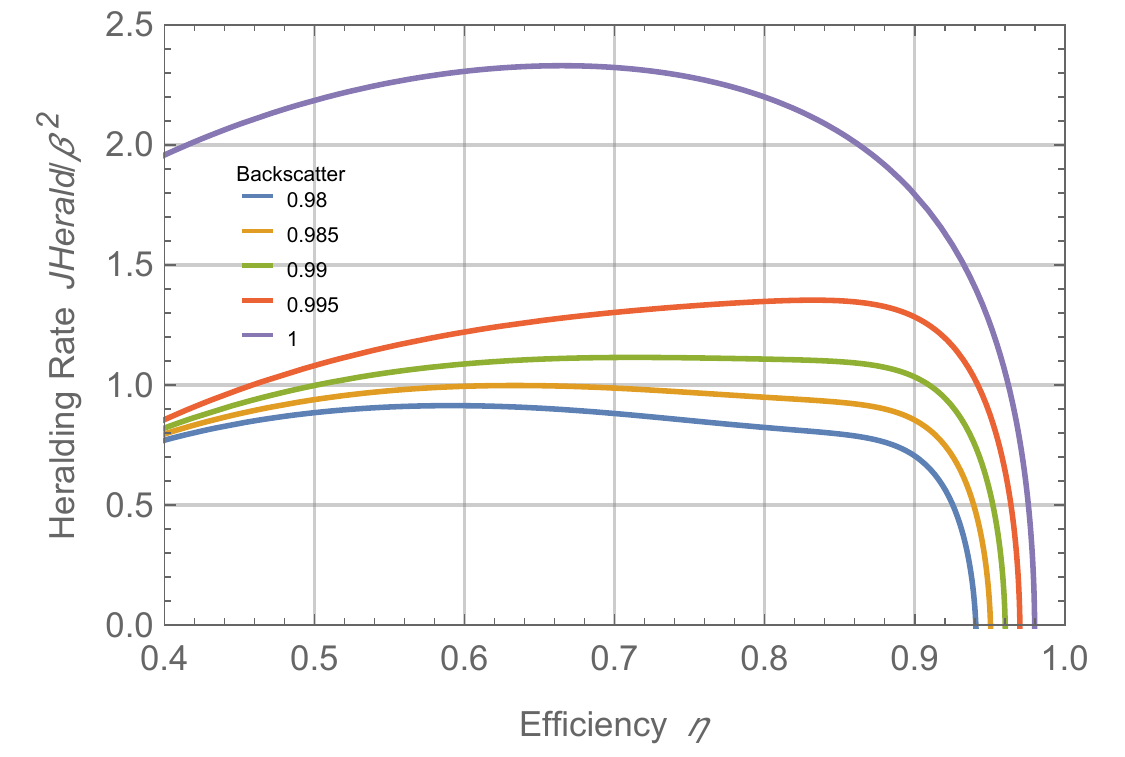}
    \caption{The logarithm (base-10) of the reduced heralding rate $J_{Herald}/\beta^2$ ($y$-axis), plotted against the heralding efficiency $\eta$ ($y$-axis), with loss coefficient $\alpha$ of 0.98. This is given for backscatter coefficient $\xi$ of magnitude 0.98 (blue), 0.985 (orange), 0.99 (green), 0.995 (red) and 1 (no backscatter, purple), and phase $\zeta$ of $0$.}
    \label{fig:ratevseffic}
\end{figure}

Attempting this numerically, in Fig. \ref{fig:ratevseffic} we get a plot of reduced heralding rate $J_{Herald}/\beta^2$ ($y$-axis) against heralding efficiency ($x$-axis). While similar to that in \cite{Vernon2016NoFree}, it shows the differences both between our model and theirs, and cases with and without backscatter. However, it does still support their conclusion - that there is a trade-off between heralding efficiency $\eta$ and heralding rate $J_{Herald}$, while also further suggesting this trade-off becomes more pronounced the greater the splitting  $\sqrt{1-\xi^2}$ becomes. It also shows that, given the position of this optimal heralding rate with respect to efficiency, that critical coupling isn't optimal for either - for both, increasing bus-ring coupling above loss (reducing $t$ to below $\alpha$) is beneficial.
 
\section{Conclusion}

We presented two models for backscatter in MRRs -  occurring in-ring, and in-coupler. We showed only the in-coupler model allows us to have asymmetrically split peaks, suggesting this in-coupler coupling must be the cause of that seen experimentally (as opposed to that caused in-ring---e.g. by wall surface roughness).

While this modelling was to allow analytic analysis of backscatter effects on four-wave mixing in these structures, they can be applied wherever MRRs are used (e.g. frequency combs, wavelength-filtering, and modulation of non-linear optical effects)
\cite{Preble2015Interference,Pasquazi2018Comb}.
Also, this analysis could be abstracted to model backscatter in any resonant cavity - which, despite being one of the key sources of difficulty in controlling their use, hasn't been heavily investigated.

Alongside this, we suggested a spectrum could act as though entirely un-split given a large enough phase, due to the split peak asymmetry this gives. This could potentially mitigate the negative effects of backscatter.

Further, we calculated the effects of backscatter on spontaneous four-wave mixing rates, heralding rates and efficiencies. A future direction would be to link these parameters to ring design, so these values could be optimised for given material parameters, to mitigate the effects of backscatter. Given how essential such sources will be for any form of optical quantum computing or quantum communication, this research will revolutionise the efficiency of these processes.

\textit{Acknowledgements} - We thank Will McCutcheon, Henkjan Gersen and Joshua Silverstone for useful discussions. This work was supported by the Engineering and Physical Sciences Research Council (Grants EP/T001011/1, EP/R513386/1, EP/M013472/1 and EP/L024020/1).

\bibliographystyle{unsrt}
\bibliography{ref.bib}

\end{document}